\documentclass[pra, aps, twocolumn, groupedaddress, showpacs, superscriptaddress]{revtex4-2}
\usepackage{amssymb,mathrsfs,amsmath,amsthm,color,graphicx,times,graphicx}
\usepackage{hyperref}
\usepackage{subfigure}
\usepackage{times}
\usepackage{bbold}
\usepackage{color}

\providecommand{\openone}{\leavevmode\hbox{\small1\kern-4.3pt\normalsize1}}

\theoremstyle{plain}

\usepackage{orcidlink}
\theoremstyle{definition}

\begin{document}
\title{Complementarity between quantum entanglement, geometrical and dynamical appearances in N spin-1/2 system under all-range Ising model}
\author{Jamal Elfakir}\affiliation{LPHE-Modeling and Simulation, Faculty of Sciences, Mohammed V University in Rabat, Rabat, Morocco.}
\author{Brahim Amghar}\email{amghar.b@ucd.ac.ma}\affiliation{Laboratory LPNAMME, Laser Physics Group, Department of Physics, Faculty of Sciences, Chouaïb Doukkali University, El Jadida, Morocco.}\affiliation{Centre of Physics and Mathematics, CPM, Faculty of Sciences, Mohammed V University in Rabat, Rabat, Morocco.}
\author{Abdallah Slaoui \orcidlink{0000-0002-5284-3240}}\email{abdallah.slaoui@um5s.net.ma}\affiliation{LPHE-Modeling and Simulation, Faculty of Sciences, Mohammed V University in Rabat, Rabat, Morocco.}\affiliation{Centre of Physics and Mathematics, CPM, Faculty of Sciences, Mohammed V University in Rabat, Rabat, Morocco.}
\author{Mohammed Daoud}\affiliation{Department of Physics, Faculty of Sciences, University Ibn Tofail, Kenitra, Morocco.}

\begin{abstract}
With the growth of geometric science, including the methods of exploring the world of information by means of modern geometry, there has always been a mysterious and fascinating ambiguous link between geometric, topological and dynamical characteristics with quantum entanglement. Since geometry studies the interrelations between elements such as distance and curvature, it provides the information sciences with powerful structures that yield practically useful and understandable descriptions of integrable quantum systems. We explore here these structures in a physical system of $N$ interaction spin-$1/2$  under all-range Ising model. By performing the system dynamics, we determine the Fubini-Study metric defining the relevant quantum state space. Applying Gaussian curvature within the scope of the Gauss-Bonnet theorem, we proved that the dynamics happens on a closed two-dimensional manifold having both a dumbbell-shape structure and a spherical topology. The geometric and topological phases appearing during the system evolution processes are sufficiently discussed. Subsequently, we resolve the quantum brachistochrone problem by achieving the time-optimal evolution. By restricting the whole system to a two spin-$1/2$ system, we investigate the relevant entanglement from two viewpoints; The first is of geometric nature and explores how the entanglement level affects derived geometric structures such as the Fubini-Study metric, the Gaussian curvature, and the geometric phase. The second is of dynamic nature and addresses the entanglement effect on the evolution speed and the related Fubini-Study distance. Further, depending on the degree of entanglement, we resolve the quantum brachistochrone problem.
\par
\vspace{0.25cm}

\end{abstract}
\date{\today}

\maketitle
\section{Introduction}
Over the past few years, there has been growing excitement about the application of geometric ideas to quantum physics. It is argued that the geometrization of quantum theory provides a significant framework able to describe, to a large extent, the physical characteristics of solvable quantum systems \cite{Kibble1979,Anandan1991,Ashtekar1999,Brody2001}. This geometrical approach has introduced the concept of the quantum phase space, endowed naturally with the Kähler manifold structure, on which the dynamics of quantum systems is well established \cite{Provost1980,Zhang1995,Botero2003,Heydari2015}. Lately, numerous studies have shown the relevance of geometric structures of the physical state space for exploring the physical properties of quantum systems. Indeed, it has been demonstrated that the Fubini-Study distance traveled by a quantum system during evolution along a given curve in relevant projective Hilbert space is related to the integral of the energy uncertainty, which in turn is proportional to the evolution speed \cite{Anandan1990}. The quantum speed limit time, which defines the fundamental limit on the rate of evolution of quantum systems, is also determined by means of Bures length between mixed quantum states \cite{Deffner2013}. Additionally, the geometric methods simplify considerably the resolution of the quantum brachistochrone problem, which is linked to determining the Hamiltonian that generates the time-optimal evolution between two states \cite{Ammghar2020,Bender2007,Frydryszak2008}. The efficient quantum circuits in quantum computation with $n$ qutrits are investigated by use of Riemannian geometry. Indeed, it has been proven that the optimal quantum circuits are basically equivalent  to the shortest path between two points in a specific curved geometry of SU($3^n$) \cite{Li2013}, analogous to the qubit case wherein the geodesic in SU($2^n$) is involved \cite{Nielsen2006}.  For other further dynamic properties explored on the basis of geometric approaches, we recommend that readers look at the papers \cite{Dowling2008,Deffner2010,Pires2015,Pires2016}.\par

Currently, the geometric quantum mechanics, which forms the theoretical framework of the geometric formulation of quantum theory, is the bedrock of  information geometric science, in which the quantum phenomena are handled geometrically in the space of quantum states. One can cite, for instance, the quantum entanglement being an intriguing physical resource in the protocols of quantum information theory \cite{Horodecki2009,Elfakir2023,Kirdi2023,Amico2008,Amghar2020}. Importantly, entanglement is shown to be closely related to the Mannoury-Fubini-Study distance separating the entangled state and the nearest disentangled state \cite{Levay2004}. Moreover, the Euclidean distance of an entangled state to the disentangled states is equal to the maximal violation of a generalized Bell inequality with the tangent functional as entanglement witness \cite{Bertlmann2002}. The connection between quantum entanglement and the state space geometry has been thoroughly explored for a spin-s system with long-range Ising interaction \cite{Krynytskyi2019}. Further to that, the geometrical description of entanglement is also explored within the backdrop of the Hopf fibration, which is a topological map compactifying the related quantum state space to an another lower-dimensional space referred to as the Hopf bundle \cite{Mosseri2006,Mosseri2001,Amghar2021}. For additional findings highlighting the interplay between quantum entanglement and geometrical characteristics, see, e.g., Refs.\cite{Amghar2023,Verstraete2002,Ha2013,Avron2009}.\par

Another significant concept that has received much attention in quantum physics is the geometric phase, a remarkable geometric characteristic in quantum evolution processes \cite{Berry1984,Aharonov1987,Anandan1992,Carollo2003}. It can be viewed as the holonomy acquired by the state vector after a parallel transport along the evolution trajectory \cite{Andersson2019,Demler1999}. The geometric phase is now intimately related to other geometrical features that define the quantum state spaces. In effect, it has been proven that the geometric phase can be expressed as the line integral of Berry-Simon connection along the Fubini-Study distance separating the two quantum states over the corresponding projective Hilbert space \cite{Botero2003,Samuel1988}. On the practical side, several recent investigations have shown the most important role of the geometric phase in the advancement of quantum information science. Indeed, it is a valuable feature for generating quantum logic gates that are critical to quantum computation \cite{Wang2001,Kleipler2018,Kumar2005}. Furthermore, the conditional phase gate has been experimentally demonstrated for nuclear magnetic resonance \cite{Jones2000} and trapped ions \cite{Duan2001}. The interplay between quantum entanglement and topological and geometric phases is also extensively studied in the two-qudit systems \cite{Oxman2011,Matoso2016,Khoury2014}. Other geometric phase applications have been realized in Refs.\cite{Johansson2012,Oxman2018,Vedral2003,Huang2015}.\par

The main purpose of this work is to exploit the geometrical structures characterizing the quantum phase space of  $N$ interacting spin-$1/2$ under all-range Ising model to explore some physical properties, such as the geometric phase, the curvature of quantum phase space, the evolution speed, distance geodesic between quantum states, solving quantum brachistochrone problem, and their interplay with quantum entanglement. It is noteworthy that the ideas explored in this paper were primarily inspired by the findings obtained by Krynytskyi and Kuzmak in Ref.\cite{Krynytskyi2019}. As a matter of fact,  by performing the system dynamics, we derive the Fubini-Study metric identifying the associated quantum state space. Moreover, examining the Gaussian curvature (G-curvature) in the framework of the Gauss-Bonnet theorem, we determine the topology and the structure of this space. Afterward, we explore the acquired geometric and topological phases and tackle the quantum brachistochrone issue. Finally, we give a detailed explanation of the geometrical and dynamical characteristics of two interacting spin-$1/2$ under the Ising model in connection to quantum entanglement.\par

The rest of this paper is structured as follows. In Sec.\ref{sec2}, by carrying out the dynamics of $N$ interacting spin-$1/2$ under all-range Ising model, we give the Fubini-Study metric and identify the associated quantum state space. Moreover, by investigating the G-curvature within the scope of the Gauss-Bonnet theorem, we uncover the topology and the structure of this space. The geometric and topological phases emerging from the system evolution processes, over the resulting state space, are thoroughly discussed in Sec.\ref{sec3}. The quantum brachistochrone problem is also addressed based on the evolution velocity as well as the related Fubini-Study distance, in Sec.\ref{sec4}. In Sec.\ref{sec5}, we study the entanglement between two interacting spin-1/2 under the Ising model from two different appearances:  the first is of a geometric nature and investigates how the entanglement degree impacts derived geometric features such as the Fubini-Study metric, the G-curvature, and the geometric phase. The second is of a dynamic type and addresses the entanglement effect on evolution speed and the corresponding Fubini-Study distance. Further to this, we resolve the quantum brachistochrone problem based on quantum entanglement. We supply concluding remarks in Sec.\ref{sec6}.
\section{Unitary evolution and the quantum state manifold of $N$ spin-$1/2$ system}\label{sec2}

\subsection{Physical model and unitary quantum evolution} 
To start, the considered system is composed of $N$ qubits represented by $N$ interacting spin-$1/2$ under all-range Ising model described by the following Hamiltonian
\begin{equation}\label{1}
 	\mathrm{H}=\mathtt{J}\left(\sum_{i=1}^{N} S^z_i\right)^{2},
 \end{equation} 
with $\mathtt{J}$ is the coupling constant characterizing the interaction and $\mathtt{S}_k^z$ denotes the $z$-component of the spin operator $\textbf{S}_i=(\mathtt{S}_i^x,\mathtt{S}_i^y,\mathtt{S}_i^z)^T$ associated with $i$th spin-$1/2$ (i.e., $i$th qubit) which fulfills the eigenvalues equation
\begin{equation}\label{2}
\mathtt{S}_i^z\left| {{\mathsf{m}_i}} \right\rangle  ={\mathsf{m}_i} \left| {{\mathsf{m}_i}} \right\rangle,
\end{equation}
where $\mathtt{S}_i^\alpha=\frac{\hbar}{2}\sigma_i^\alpha$ and $\sigma_i^\alpha$, $(\alpha=x,y,z)$ are the Pauli matrices, $\mathsf{m}_{i}=\pm \hbar/2$ represent the possible values due to the projection of the $ith$ spin over the $z$-axis, and $\left| {{\mathsf{m}_i}} \right\rangle $ denote the associated eigenstates. It is worth noting that the components of spin-$1/2$ operators $\mathtt{S}_i^x,\,\mathtt{S}_i^y,$ and $\mathtt{S}_i^z$ satisfy the algebraic structure of the Lie group SU(2):
\begin{equation}
\left[ {\mathtt{S}_i^\alpha ,\mathtt{S}_j^\beta } \right] = i{\delta _{ij}}\sum\limits_{\gamma  = x,y,z}\epsilon^{\alpha\beta\gamma} {\mathtt{S}_i^\gamma },
\end{equation}
where $\delta _{ij}$ and $\epsilon^{ijk}$ designate the Kronecker and Levi-Civita symbols, respectively. It is straightforward to see that for an even number of spins, the above Hamiltonian has $(N/2 + 1)$ eigenvalues, whereas for an odd number, it has $(N+ 1)/2$ eigenvalues. Explicitly, the eigenvalues and related eigenstates are provided as follows
	\begin{equation}\label{3}
\begin{array}{cc}
\frac{\mathtt{J}\hbar^2}{4} N^2 & |\uparrow \uparrow \uparrow \ldots \uparrow \uparrow\rangle,|\downarrow \downarrow \downarrow \ldots \downarrow \downarrow\rangle ; \\[12px]
\frac{\mathtt{J}\hbar^2}{4}(N-2)^2 & |\downarrow \uparrow \uparrow \ldots \uparrow \uparrow\rangle,|\uparrow \downarrow \uparrow \ldots \uparrow \uparrow\rangle, \ldots,|\uparrow \uparrow \uparrow \ldots \uparrow \downarrow\rangle, \\[6px]
& |\uparrow \downarrow \downarrow \ldots \downarrow \downarrow\rangle,|\downarrow \uparrow \downarrow \ldots \downarrow \downarrow\rangle, \ldots,|\downarrow \downarrow \downarrow \ldots \downarrow \uparrow\rangle ; \\[12px]
\frac{\mathtt{J}\hbar^2}{4}(N-4)^2 & |\downarrow \downarrow \uparrow \ldots \uparrow \uparrow\rangle,|\downarrow \uparrow \downarrow \ldots \uparrow \uparrow\rangle, \ldots,|\uparrow \uparrow \uparrow \ldots \downarrow \downarrow\rangle, \\[6px]
& |\uparrow \uparrow \downarrow \ldots \downarrow \downarrow\rangle,|\uparrow \downarrow \uparrow \ldots \downarrow \downarrow\rangle, \ldots,|\downarrow \downarrow \downarrow \ldots \uparrow \uparrow\rangle ; \\[12px]
\ldots & \ldots
\end{array}
	\end{equation}
	Taking into account all possible combinations of spin states (i.e., $|\uparrow\rangle$ and $|\downarrow\rangle$), one finds that each eigenvalue $
{{J\hbar^2{{(N - 2p)}^2}} \mathord{\left/
 {\vphantom {{J{{(N - 2p)}^2}} 4}} \right.
 \kern-\nulldelimiterspace} 4}$ matches $2\mathrm{C}_N^p$ eigenstates, with $\mathrm{C}$ denotes for the binomial coefficient, while the index $p=0,...,N/2$ for $N$ even (particle number)  and $p=0,...,(N-1)/{2}$ for $N$ odd. We presume that the evolution of the $N$ spin-$1/2$ system starts with the initial state
 \begin{equation}\label{a}
 |\Psi_i\rangle=|+\rangle^{\otimes N},
 \end{equation}
		where
\begin{equation*}
|+\rangle=\cos\frac{\theta}{2}|\uparrow\rangle+\sin\frac{\theta}{2}e^{i\varphi}|\downarrow\rangle,
\end{equation*}		
corresponds to the eigenstate of the spin-$1/2$ projection operator on the direction denoted by the unit vector $\textbf{n}=(\sin\theta\cos\varphi,\sin\theta\sin\varphi,\cos\theta)$, where $\theta$ and $\varphi$ designate the polar and azimuthal angles, respectively. In this respect, the initial state \eqref{a} of the system can be rewritten, using the binomial theorem, as
\begin{small}
\begin{equation} \label{b}
\left|\Psi_i\right\rangle=\sum_{p=0}^N \cos^{N-p}\frac{\theta}{2}\sin^p\frac{\theta}{2}e^{ip\varphi} \sum_{i_1<i_2<\ldots<i_p=1}^N \sigma_{i_1}^x \sigma_{i_2}^x \ldots \sigma_{i_p}^x|\uparrow \rangle^{\otimes N},
		\end{equation}
\end{small}
where we set $\hbar= 1$, indicating that the energy is measured in the frequency units. To investigate the geometrical, topological, and dynamical features of the system under consideration in the remainder of this paper, we need to evolve the $N$ spin-1/2 system, maintained initially in the starting state \eqref{b}, by applying the time evolution propagator $\mathcal{P}(t)=e^{-i\mathrm{H}t}$. In this perspective, the evolving state of the system is obtained as
		\begin{widetext}
		\begin{equation}\label{c}
			\left|\Psi(t)\right\rangle=\sum_{p=0}^N \cos^{N-p}\left( \frac{\theta}{2}\right)\sin^p\left(\frac{\theta}{2}\right)  \exp\left\lbrace-i\left[\frac{\xi(t)}{4}(N-2p)^2-p\varphi\right]\right\rbrace \sum_{i_1<i_2<\ldots<i_p=1}^{N} \sigma_{i_1}^x \sigma_{i_2}^x \ldots \sigma_{i_p}^x|\uparrow \uparrow \ldots \uparrow\rangle,
		\end{equation}
		\end{widetext}
where we define $\xi(t)=\mathtt{J}t$ as the dynamical parameter of the system. In this way, we come at demonstrating the evolved state of a collection of $N$ qubits (i.e.,  $N$ spin$-1/2$) that is explicitly dependent on three parameters, namely the spherical angles $(\theta,\varphi)$ and the dynamical parameter $\xi$. It is also intriguing to note that the state \eqref{c} fulfills the periodic requirement $\left|\Psi(\xi)\right\rangle =\left|\Psi(\xi+2\pi)\right\rangle$ along the temporal parameter, implying that $\xi\in[0,2\pi]$. Accordingly, one can predict that the dynamics of the system occurs on a closed three-dimensional manifold.
 \subsection{Geometry and topology of the resulting state manifold}
 After evolving the $N$ spin-$1/2$ system by means of evolution operator and  determining the evolved state \eqref{c}, we will now discover the geometry and topology associated with the relevant quantum state space, which includes all possible states that the system may reach during evolution. For this task, we must compute the Fubini-Study metric, which defines the infinitesimal distance $d\mathtt{S}$ between two adjoining pure quantum states $|\Psi (\zeta^\mu)\rangle$ and $|\Psi (\zeta^\mu+d\zeta^\mu)\rangle$, having the following form \cite{Tkachuk2011,Abe1993}
 		\begin{equation}\label{d}
 	d\mathtt{S}^2=\mathrm{g}_{\mu\nu} d \zeta^\mu d \zeta^\nu,	
 		\end{equation}
where $\zeta^\mu$ are the physical parameters $\theta, \varphi$ and $\xi$ specifying the evolved state \eqref{c} and $\mathrm{g}_{\mu \nu}$ denote the components of this metric tensor given by \cite{Tkachuk2011}
\begin{equation}\label{e}
\mathrm{g}_{\mu \nu}=\mathrm{Re}\left(\left\langle\Psi_\mu|\Psi_\nu\right\rangle-\left\langle\Psi_\mu |\Psi\right\rangle\left\langle\Psi| \Psi_\nu\right\rangle\right),
\end{equation}
with $\left|\Psi_{\mu}\right\rangle=\frac{\partial}{\partial \zeta^{\mu}}|\Psi\rangle$. Using the definition \eqref{d}, one finds, after a simple numerical calculation, the explicit version of the Fubini-Study metric as follows
\begin{small}
\begin{align}\label{e}
 	d\mathtt{S}^2=& d\mathtt{S}_i^2+ \frac{1}{4} N(N-1) \sin^2 \theta\left[N-1-\left(N-\frac{3}{2}\right) \sin^2 \theta\right]d\xi^2\notag\\& + \frac{1}{4} N(N-1)\cos \theta \sin ^2 \theta d\varphi d\xi,
 		\end{align}
\end{small}	
with
\begin{small}
\begin{equation}\label{a17}
d\mathtt{S}_i^2= \frac{N}{4}\left(d \theta^2+\sin^2 \theta d \varphi^2\right),
\end{equation}
\end{small}
corresponds to the  squared line element defining the sphere of possible initial states of the system. This can be well remarked when we take $\xi=0$ (i.e., no evolution) in the metric \eqref{e}, we discover indeed that the state space is reduced to a sphere of radius $2\sqrt{N}$. Furthermore, the space of $N$ spin-$1/2$ states resulting from the temporal evolution is effectively a closed three-dimensional manifold. Note that the components of the underlying metric \eqref{e} are $\varphi$-independent, signifying that the quantum state spaces with a predesignated azimuthal angle have the same geometry. Hence, we draw the conclusion that the appropriate quantum state space (i.e., quantum phase space) corresponding to the $N$ qubits (under consideration) is a two-parametric and curved manifold, it is identified by the following metric tensor
\begin{small}
\begin{equation}\label{f}
 	d\mathtt{S}^2=\frac{N}{4}d \theta^2 + \frac{1}{4} N(N-1) \sin^2 \theta\left[N-1-\left(N-\frac{3}{2}\right) \sin ^2 \theta\right]d\xi^2.
 		\end{equation}
\end{small}
To further characterize this state space, we are going to determine its topology. For this aim, we begin by assessing the corresponding G-curvature, measuring the intrinsic curvature of the resulting quantum state manifold \eqref{f}. It can be defined in terms of the relevant metric tensor \eqref{f} in the form \cite{Kolodrubetz2013}
\begin{align}\label{g}
\mathrm{K}=\frac{1}{\left(\mathrm{g}_{\theta \theta} \mathrm{g}_{\xi \xi}\right)^{1 / 2}} &\left[\frac{\partial}{\partial \xi}\left(\left(\frac{\mathrm{g}_{\xi \xi}}{\mathrm{g}_{\theta \theta}}\right)^{1 / 2} \Gamma_{\theta \theta}^{\xi}\right)\right. \notag\\&\left.-\frac{\partial}{\partial \theta}\left(\left(\frac{\mathrm{g}_{\xi \xi}}{\mathrm{g}_{\theta \theta}}\right)^{1 / 2} \Gamma_{\theta \xi}^{\xi}\right)\right],
\end{align}
where   $\Gamma_{\theta \theta}^{\xi}$ and $\Gamma_{\theta \xi}^{\xi}$ account for the Christoffel symbols given by
\begin{equation}
\Gamma_{\theta \theta}^{\xi}=-\frac{1}{2 \mathrm{g}_{\xi \xi}}\left(\frac{\partial \mathrm{g}_{\theta \theta}}{\partial \xi}\right), \quad \text{and}  \quad \Gamma_{\theta \xi}^{\xi}=\frac{1}{2 \mathrm{g}_{\xi \xi}}\left(\frac{\partial \mathrm{g}_{\xi \xi}}{\partial \theta}\right).
\end{equation}
It is extremely intriguing to see that the temporal component $\mathrm{g}_{\xi \xi}$ of the metric cancels out at the points $\theta=0, \pi$. This implies that the G-curvature is not definable in these positions, hence we conclude that it exhibits a singularity in each of these two positions. However, it is well definable at all other positions within the space of $N$ spin-1/2 states. Reporting the metric components $\mathrm{g}_{\theta \theta}$ and $\mathrm{g}_{\xi \xi}$ in the equation \eqref{g}, the explicit expression of the relevant G-curvature writes
\begin{equation} \label{h}
 			\mathrm{K}=\frac{8}{N}\left[2-\frac{(2N-3) \cos^2 \theta+N}{\left((2 N-3) \cos ^2 \theta+1\right)^2}\right].
 		\end{equation}
Note that the state space curvature \eqref{h} is mainly affected by the initial parameters $\theta$ and $N$, while it is independent of $\xi$ containing the temporal evolution, meaning that the state space curvature is independent of the system dynamics. Furthermore, the G-curvature \eqref{h} verifies the following periodic requirement $\mathrm{K}(\theta)=\mathrm{K}(\theta+\pi)$. This is consistent with our findings because the resulting quantum phase space \eqref{f} is a closed two-dimensional manifold.
 \begin{figure}[h]
\begin{center}
\includegraphics[scale=0.48]{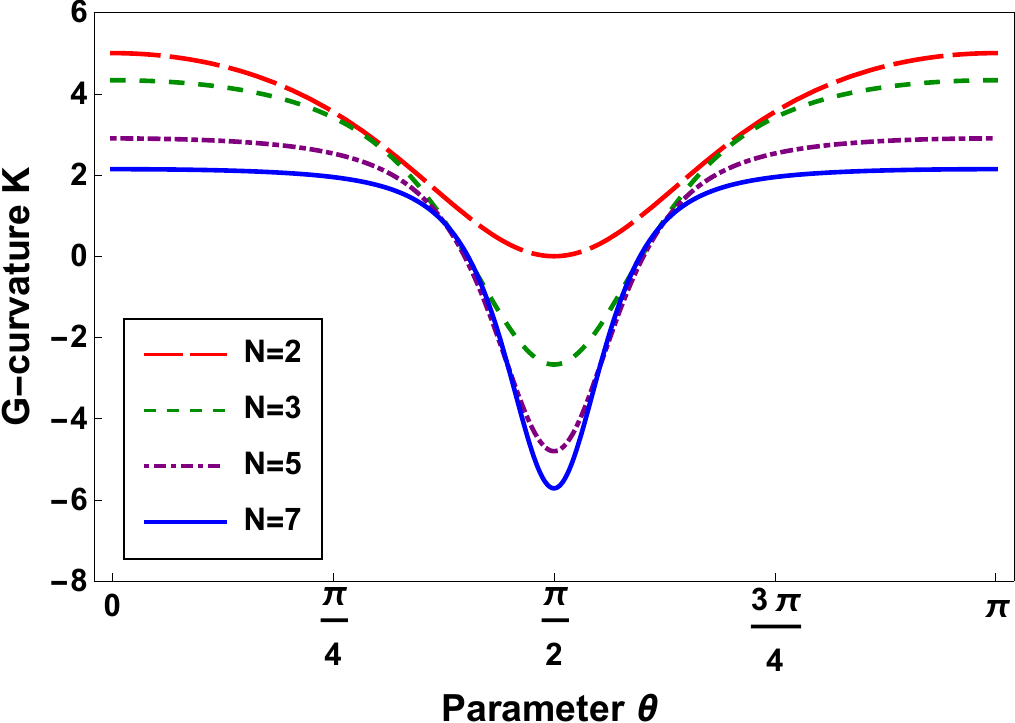}
\caption{The dependence of the G-curvature \eqref{h} on the initial parameter $\theta$ for some spin-$1/2$ numbers.}\label{am}
\end{center}
\end{figure}

From the analysis of the Fig.\eqref{am} showing the G-curvature behavior with respect to the initial parameters $(\theta,N)$, we see that the state space geometry is symmetric with respect to the centerline at $\theta=\pi/2$, being the position with the minimal curvature. More specifically, in the region $\theta\in[0,\pi/2]$, the G-curvature declines and therefore the state space geometry takes the concave shape, whereas in the region $\theta\in[\pi/2,\pi]$, the G-curvature increases to its maximum value, and therefore the space geometry takes the convex shape. As a result, we infer that the $N$ spin-$1/2$ state space has a dumbbell-shape structure. Additionally, we notice that for $N>2$, the curvature takes negative values  for certain values of $\theta$. This is congruent with the findings given in Ref. \cite{Krynytskyi2019}.\par
Considering this fact, in addition to the existence of two singularities in the G-curvature \eqref{h}, we get to the conclusion that there are two conical flaws within the quantum phase space \eqref{f}; the first one is situated near to the location $\theta= 0$, whereas the second one is situated near to the location $\theta=\pi$. In light of these outcomes, let us now explore the topology associated with the space of $N$ spin-$1/2$ states \eqref{f}. To achieve this, we have to compute the integer Euler characteristic $\chi(\mathrm{M})$ ($\mathrm{M}$ stands for state space \eqref{f}) provided in the Gauss-Bonnet theorem as \cite{Kolodrubetz2013}
 		\begin{equation}\label{i}
\frac{1}{2 \pi}\left[\int_{\mathrm{M}} \mathrm{K} d\mathrm{S}+\oint_{\partial \mathrm{M}} \mathrm{k}_{\mathrm{g}} d{l}\right]=\chi(\mathrm{M}),
 		\end{equation}
where the geometric invariants $d\mathrm{S}$, $\mathrm{k}_{\mathrm{g}}$ and $d{l}$ denote, respectively, the surface element, geodesic curvature, i.e., the curvature at the boundary of the state space \eqref{f}, and line element. Additionally, the first and second terms on the left-hand side of the equation \eqref{i} represent, respectively, the bulk and border contributions to the Euler characteristic identifying the state space topology. Furthermore, the Gauss-Bonnet theorem \eqref{i} can be established in terms of the state space geometry \eqref{f} in the form
 \begin{equation}\label{j}
 \int_0^\pi \int_0^{{2\pi }} \mathrm{K}\left(\mathrm{g}_{\theta \theta} \mathrm{g}_{\xi \xi}\right)^{1 / 2} d \theta d \xi+{\Lambda}=2 \pi \chi(\mathrm{M}),
\end{equation} 		
here $\Lambda$ denotes the Euler border integral including the conical flaws contribution. After a simple calculation, the Gauss-Bonnet theorem \eqref{j} reads as
\begin{equation}\label{l}
4 \pi(N-1)+\Lambda=2 \pi \chi(\mathrm{M}).
\end{equation}
Therefore, to find the Euler characteristic $\chi$, we must first determine the Euler border integral $\Lambda$. For this purpose, we presume that the angular flaws are situated very near to the singular points $\theta=0, \pi$. In this view, the underlying metric \eqref{f} can be evaluated, in the vicinity of these two singular positions, upto the second order in $\theta$. Indeed, we obtain
 	\begin{equation}\label{k01}
 		d\mathtt{S}^2=\frac{N}{4}d\theta^2+\frac{1}{4}N(N-1)^2\theta^2 d\xi^2.
 		\end{equation}	
 It follows that
 		\begin{equation}\label{k}
 		\Lambda=2\left[2 \pi-\frac{2\pi\sqrt{\mathrm{g}_{\xi \xi}} }{\sqrt{\mathrm{g}_{\theta \theta}} \theta}\right]=4\pi\left(2 -N\right),
 		\end{equation}
where we multiplied it by the factor 2 because we have two singular points. The detailed derivation of the result \eqref{k} is provided in the appendix. Putting the equation \eqref{k} into \eqref{l}, one finds the Euler characteristic $\chi(\mathrm{M})=2$, showing that the quantum phase space associated with the $N$ spin-$1/2$ system has a spherical topology. The coming section will be devoted  to a thorough analysis of the geometric phase that the $N$ spin-$1/2$ state \eqref{c} can accumulate when subjected to cyclic and arbitrary evolution processes on the underlying quantum state space \eqref{f}.
 \section{Geometrical phases acquired by the $N$ spin-$1/2$ state}\label{sec3}
 After investigating the geometry and topology of the $N$ spin-$1/2$ state space identified by the metric tensor \eqref{f}, let us now focus on the geometric phase that the evolving state \eqref{c} can acquire for both arbitrary and cyclic evolutionary processes.
 \subsection{Geometrical phase during an arbitrary evolution}
In this instance, we presume that the $N$ spin-$1/2$ system evolves arbitrarily along any evolution path on the closed two-dimensional manifold \eqref{f}. In this picture, the geometric phase gained by the evolved state \eqref{c} is given by
\begin{equation}\label{m}
\Phi_g(t)=\arg \langle\Psi_i| \Psi(t)\rangle-{\mathrm{Im}} \int_0^t\langle\Psi(t^{\prime})|\frac{\partial}{\partial t^{\prime}}| \Psi(t^{\prime})\rangle d t^{\prime},
\end{equation}
which is defined as the difference between total phase and dynamic phase \cite{Oxman2011,Mukunda1993}. To calculate the geometric phase, we must first compute total the phase acquired by the system, The overlap (i.e., transition-probability amplitude) between the starting state \eqref{b} and the ending state \eqref{c} is obtained as
\begin{equation}\label{n}
\langle\Psi_i| \Psi(t)\rangle=\sum_{p=0}^{N}\mathrm{C}_N^p\cos^{2(N-p)}
\left( \frac{\theta}{2}\right)
\sin^{2p}
\left(\frac{\theta}{2}\right)e^{-\frac{i\xi}{4}(N-2p)^2}.
\end{equation}
Inserting the expression of the overlap \eqref{n} into the first term on the right side of the equation \eqref{m}, the total phase gained by the $N$ spin-$1/2$ system writes
\begin{small}
\begin{equation}\label{o}
 			\Phi_{\operatorname{tot}}=-\arctan\left[
 			\frac{\sum\limits_{p=0}^{N}\mathrm{C}_N^p\cos^{2(N-p)}\frac{\theta}{2}\sin^{2p}\frac{\theta}{2}\sin\left(\frac{\xi(N-2p)^2}{4}\right)}{\sum\limits_{p=0}^{N}\mathrm{C}_N^p\cos^{2(N-p)}\frac{\theta}{2}\sin^{2p}\frac{\theta}{2}\cos\left(\frac{\xi(N-2p)^2}{4}\right)}
 			\right].
\end{equation}
\end{small}
It is interesting to observe that the total phase \eqref{o} comprises two distinct phase components: the first is of geometrical origin (known as the geometrical phase), and it is strongly related to the geometrical and topological features that characterize the quantum state space of such systems \cite{Oxman2011,Kolodrubetz2013}. This geometric component is explained by the implicit reliance of the total phase \eqref{o} on the G-curvature \eqref{h} and the component $\mathrm{g}_{\xi \xi}$ of the metric \eqref{f}, as they all depend on the parameters $(N,\theta)$. The second one is of dynamical origin (known as the dynamical phase), and it results from the time evolution of Hamiltonian eigenstates \eqref{3}. Furthermore, the global phase \eqref{o} exhibits a non-linear time dependence and fulfills the following periodic conditions :
\begin{equation}
\Phi_{\operatorname{tot}}(\xi+4\pi)=\Phi_{\operatorname{tot}}(\xi)\qquad \text{for}\; N \; \text{integer},
\end{equation}
and
\begin{equation}
\qquad\Phi_{\operatorname{tot}}(\xi+8\pi)=\Phi_{\operatorname{tot}}(\xi)\qquad \text{for}\; N \; \text{half-integer}.
\end{equation}
The dynamical phase, on the other hand, can be calculated by plugging the evolved state \eqref{c} into the second term on the right side in the equation \eqref{m}. Indeed, one finds 
\begin{equation}\label{u}
\Phi_{\operatorname{dyn}}=-\frac{\xi N}{4}\left(N\cos ^2 \theta+\sin ^2 \theta\right).
\end{equation}
It is proportional to the evolution time, meaning the dynamic phase primarily instructs us about the time spent by the system during evolution. On the other hand, the geometric phase that can be accrued by the $N$ spin-$1/2$ state \eqref{a}, during any arbitrary evolution over the quantum phase space \eqref{f}, is obtained as
\begin{small}
\begin{align}\label{p}
\Phi_{\operatorname{g}}=& -\arctan\left[
 			\frac{\sum\limits_{p=0}^{N}\mathrm{C}_N^p\cos^{2(N-p)}\frac{\theta}{2}\sin^{2p}\frac{\theta}{2}\sin\left(\frac{\xi(N-2p)^2}{4}\right)}{\sum\limits_{p=0}^{N}\mathrm{C}_N^p\cos^{2(N-p)}\frac{\theta}{2}\sin^{2p}\frac{\theta}{2}\cos\left(\frac{\xi(N-2p)^2}{4}\right)}
 			\right]\notag\\[5px]&+\frac{\xi N}{4}\left(N\cos ^2 \theta+\sin ^2 \theta\right).
\end{align}
\end{small}
It is clear that the resulting geometric phase \eqref{p} varies (i.e., accumulates or loses) non-linearly with the time, reflecting its dynamic character. Otherwise, it is dependent on the freedom degrees $(\theta, \xi)$ specifying the physical states over the quantum phase space \eqref{f}, which means that the geometric phase depends on the shape of the evolution trajectory followed by the system, while its reliance on the initial parameters $(N,\theta)$  signifies that it is also sensitive to the state space geometry. Accordingly, we conclude that the geometric phase \eqref{p} can be exploited to parameterize the possible evolution trajectories of this system. This result can find applications in quantum computation, because such quantum phases can be used to design logic gates that are helpful for building good quantum algorithms \cite{Kumar2005,Wang2002}. Let us now turn to a special scenario in which we investigate the geometric phase accrued by the $N$ spin-$1/2$ state \eqref{c} over a very brief period of time. In this framework, by extending the exponential factor given in \eqref{n} up to the second order in $\xi$, one obtains
\begin{widetext}
\begin{small}
\begin{equation}
  \langle \Psi_i|\Psi(t)\rangle\simeq 1+\frac{\xi^2 N(N-1)}{64}\left[
  4 (N-1)(N+2)\cos^2\theta-(N-3)(N-2)\sin^22\theta + 4(3N-2)
  \right]
  -i\frac{\xi N}{4} \left(
  N\cos^2\theta+\sin^2\theta
  \right).
  \end{equation}
  \end{small}
  In this perspective, the geometric phase \eqref{p} can be expressed as
 \begin{small} 
\begin{equation}
\Phi_{\operatorname{g}}\simeq -\arctan\left[
 			\frac{{\xi N} \left(
  N\cos^2\theta+\sin^2\theta
  \right)}{4+\frac{\xi^2 N(N-1)}{16}\left[
  4 (N-1)(N+2)\cos^2\theta-(N-3)(N-2)\sin^22\theta + 4(3N-2)
  \right]}
 			\right]+\frac{\xi N}{4}\left(N\cos^2 \theta+\sin^2 \theta\right).
\end{equation}
\end{small}
\end{widetext}
Note that for $\xi=0$, the system does not gain any quantum phase, this is because the system state remains confined in the starting state \eqref{b} (i.e. no evolution). As we can see that the greater the number of particles $N$, the more the dynamic phase is dominant. Further, in the thermodynamic limit ($N\to\infty$), the total phase cancels out and therefore the geometric and dynamic phases coincide at any moment in the evolution process. This offers us the opportunity of measuring the geometric phase experimentally because, in this situation, it can be obtained by the temporal integral of the average eigenvalues of the Ising Hamiltonian \eqref{1}.
\subsection{Geometrical phase under a cyclic evolution}
\begin{figure}[h]
\begin{center}
\includegraphics[scale=0.48]{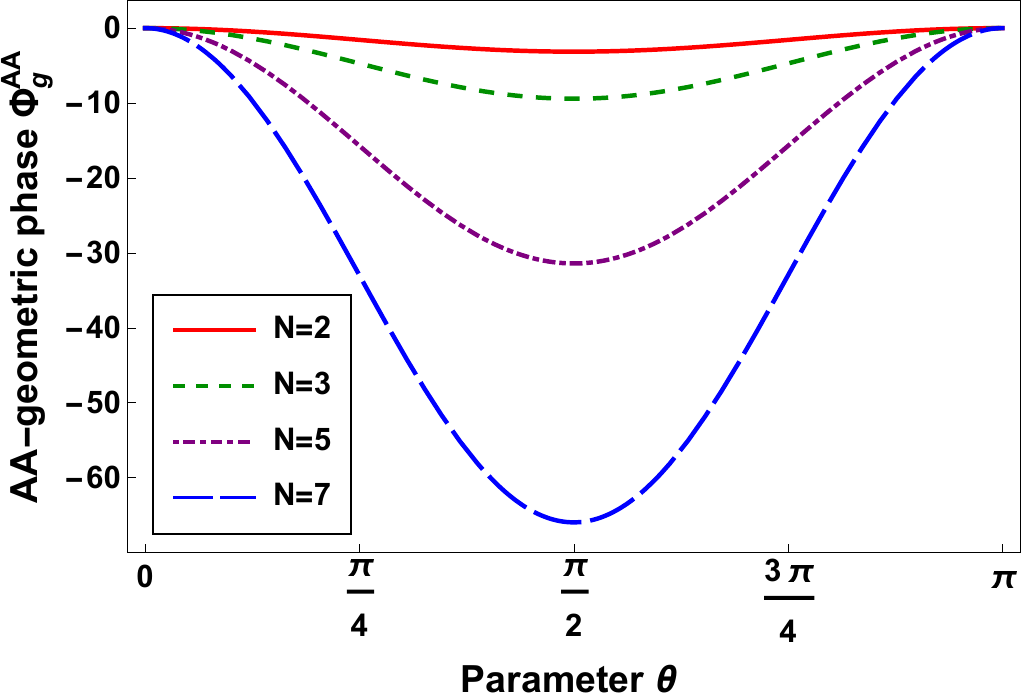}
\caption{The dependence of the AA-geometric phase \eqref{x1} on the initial parameter $\theta$ for some spin-$1/2$ numbers.}\label{av}
\end{center}
\end{figure}

Here, we are focusing on the study of the geometrical phase resulting from the cyclic evolution of the $N$ spin-$1/2$ system. In this regard, the wave function \eqref{c} satisfies the cyclic requirement $|\Psi({T})\rangle=e^{i \Phi_{\operatorname{tot}}}|\Psi(0)\rangle$ wherein ${T}$ represents the time span for a cyclic evolution. The AA-geometric phase (often referred to as the Aharonov-Anandan phase) gained by the system after a cyclic evolution (i.e., closed curve on the related parameter space) is given by \cite{Aharonov1987,Pati1995}
\begin{equation}\label{r}
\Phi_{\operatorname{g}}^{\mathrm{AA}}=i \int_0^{{T}}\langle\tilde{\Psi}(t)|\frac{\partial}{\partial t}| \tilde{\Psi}(t)\rangle d t,
\end{equation}
here $|\tilde{\Psi}(t)\rangle$ denotes the Anandan-Aharonov section given in Ref. \cite{Aharonov1987}. It is identified through the evolved state \eqref{c} as  $|\tilde{\Psi}(t)\rangle=e^{-i {f}(t)}|\Psi(t)\rangle$ where ${f}(t)$ is any smooth function satisfying $f({T})-f(0)=\Phi_{\operatorname{tot}}$. On this view, the AA-geometric phase \eqref{r} rewrites
\begin{equation}\label{w}
\Phi_{\operatorname{g}}^{\mathrm{AA}}=\int_0^{{T}} d \Phi_{\operatorname{tot}}+i \int_0^{{T}}\langle\Psi(t)|\frac{\partial}{\partial t}| \Psi(t)\rangle dt.
\end{equation}
Reporting the equations \eqref{o} and \eqref{u} into \eqref{w}, one obtains the AA-geometric phase acquired by the $N$ spin-$1/2$ system \eqref{c} during a cyclic evolution of the form
\begin{equation}\label{x1}
\Phi_{\operatorname{g}}^{\mathrm{AA}}= -\frac{\pi}{2}N(N-1)\sin^2 \theta.
\end{equation}
Thus, the obtained AA-geometric phase \eqref{x1} is independent of the system dynamics. Otherwise, it depends only on  the initial parameters $\theta$ and $N$ (i.e., the starting state), controlling the shape of the state space \eqref{f}. As a result, we conclude that the AA-geometric phase is impacted by the state space geometry and not by the evolution path followed by the system. Hence the cyclic evolution paths are not parameterizable by this cyclic phase. Further, using the equation \eqref{h} into \eqref{x1}, one can relate the AA-geometric phase with the G-curvature as follows
\begin{equation}
\Phi_{\operatorname{g}}^{\mathrm{AA}} = \frac{{\pi N(N- 1)}}{2}\left[ {\frac{{ - 56 + 3N\left( {16 - (N- 1)\mathrm{K}} \right)}}{{\left( {2N- 3} \right)\left( {N\mathrm{K} - 16} \right)}}} \right].
 			\end{equation}
 In Fig.\eqref{av}, we depict the behavior of the cyclic geometrical phase \eqref{x1} with respect to the starting parameters $(N,\theta)$. We find that it behaves similarly to G-curvature (see Figure 1).
 Specifically, the AA-geometric phase \eqref{x1} decreases in the region $\theta\in [0,\pi/2]$, in which the state space geometry has a concave shape (i.e., the G-curvature decreases), whereas it grows in the region $\theta\in [\pi/2,\pi]$, in which this geometry has a convex form (i.e., the G-curvature increases). Thus, we conclude that the AA-geometric phase \eqref{x1} also has a symmetric behavior along the parameter $\theta$, this is due to the dumbbell-shape structure of the state space. Additionally, we see that it is interesting to discuss the topological phase that can emerge during the cyclic evolution of the wave function \eqref{c}. In reality, it constitutes the part of the cyclic geometric phase that does not receive any dynamic contribution. Explicitly,  it is found as
 \begin{equation}\label{a1}
 \Phi_{\operatorname{top}}^{\mathrm{AA}} =-\frac{\pi}{2}N^2.
 \end{equation}
 The resulting topological phase \eqref{a1} is propotional to the square of the particle number. Especially, this phase takes fractional values for $N$ odd and multiples of $\pi$ for $N$ even. This shows that the topological phase (resp. the particle number) is relevant to control the state space topology given in \eqref{l}. This provides the possibility to parametrize the closed evolution paths traversed by the system through the resulting topological phase \eqref{a1}. This result looks very interesting in quantum computing, particularly in the search for efficient quantum circuits \cite{Huang2015,Roushan2014}. This issue can be closely connected to the determination of the optimal evolution path of the system under consideration, by evaluating the evolution speed as well as the related Fubini-Study distance. This is the quantum Brachistochrone problem, which will be tackled in the succeeding section.
 \section{Speed and quantum brachistochrone issue for $N$ spin-$1/2$ system}\label{sec4}
 Now, we will exploit the Riemannian geometry identifying the quantum state space \eqref{f} to investigate some dynamical properties of the system. In particular, we examine the evolution speed and the geodesic distance measured by the Fubini-Study metric \eqref{f} in order to solve the  related quantum brachistochrone problem \cite{Ammghar2020,Mostafazadeh2007}. This issue is often linked to achieving time-optimal evolution, which is characterized by a maximum speed and the trajectory between the starting state \eqref{b} and the ending state \eqref{c} is the shortest possible. In other terms, the solution to this dynamic problem is to find the shortest period of evolution.
  
\subsection{Speed of quantum evolution}
 In order to evaluate the evolution speed, we presume that the evolution of the $N$ spin-$1/2$ system depends only on time while leaving all other parameters unchanged. In this picture, the metric tensor \eqref{f} simplifies to
 \begin{equation}\label{k06}
 d\mathtt{S}^2=\mathrm{g}_{\xi \xi}d\xi^2.
 \end{equation}
 Explicitly, we obtain
 \begin{equation}\label{a4}
 	d\mathtt{S}^2=\frac{1}{4} N(N-1) \sin^2 \theta\left[N-1-\left(N-\frac{3}{2}\right) \sin^2 \theta\right]d\xi^2.
 		\end{equation}
 This shows that the dynamics of the system occurs on a circle of radius $\sqrt{\mathrm{g}_{\xi \xi}}$. Therefore, the evolution speed of the $N$ spin-$1/2$ state \eqref{c} takes the form \cite{Anandan1990}
 \begin{equation}\label{a5}
 \mathrm{V}=\frac{d \mathtt{S}}{d t}={2}\Delta \mathrm{E},
 \end{equation}
 where $\Delta \mathrm{E}$ designates the energy uncertainty of the above Hamiltonian \eqref{1}. From the analysis of the equation \eqref{a5}, we notice that the larger the energy uncertainty, the faster the system evolves, and vice versa. By setting the equation \eqref{a4} into \eqref{a5}, the evolution velocity of the wave function \eqref{c} is found as
 \begin{equation}\label{a7}
  \mathrm{V}=\frac{\mathtt{J}}{2}\sqrt{N(N-1) \sin^2 \theta\left[N-1-\left(N-\frac{3}{2}\right) \sin^2 \theta\right]}.
 \end{equation}
 Thus, the evolution rapidity is affected by the coupling interaction $\mathtt{J}$ and the initial parameters ($\theta$, $N$), i.e., the choice of the starting state. From the equation \eqref{a7}, we remark that the larger the particle number and the coupling constant, the faster the system evolves, with the exception of $\theta=0$ or $\theta=\pi$, in which the evolution velocity \eqref{a7} cancels out $(\mathrm{V} =0)$ regardless of these physical parameters. This is justified by the fact that neither the wave function \eqref{c} nor the G-curvature \eqref{g} are defined in these two singular points.\par
\begin{figure}[h]
\begin{center}
\includegraphics[scale=0.48]{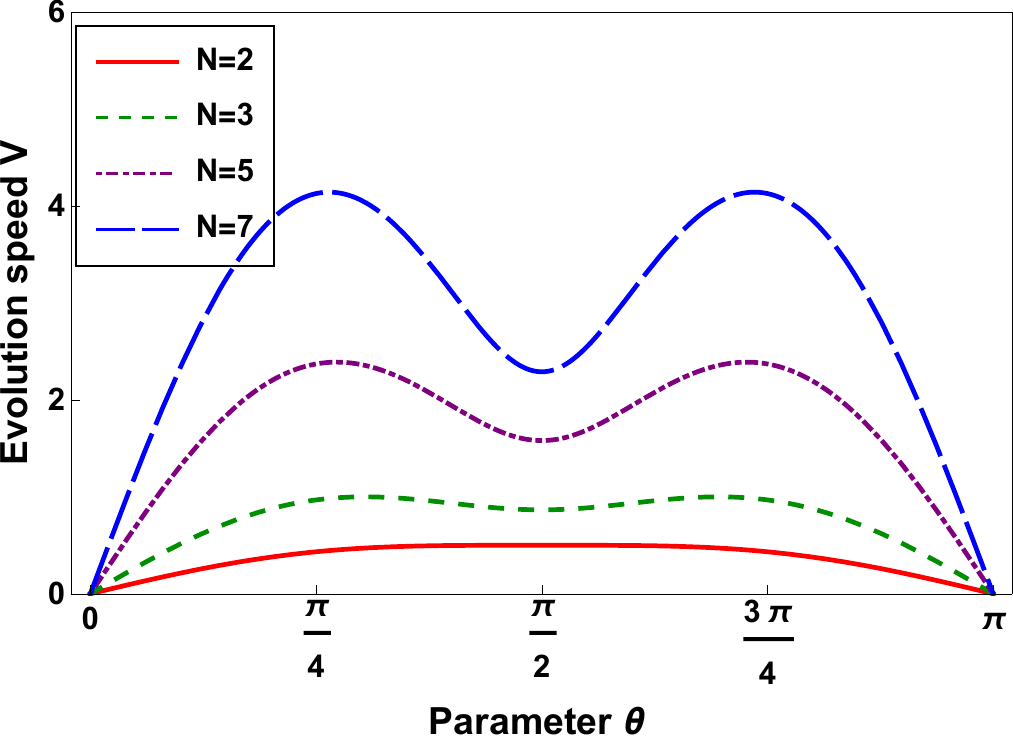}
\caption{The dependence of the evolution speed \eqref{a7} on the initial parameter $\theta$ for some spin-$1/2$ numbers with $\mathtt{J}=1$.}\label{al}
\end{center}
\end{figure}

 Given that the speed depends parameter $\theta$, meaning that it is also assigned both by the G-curvature \eqref{h} and the geometric phase \eqref{p} including the AA-geometric-phase \eqref{x1}. This can be clearly seen in the figure \eqref{al} displaying its reliance on the initial parameters $(\theta,N)$.

Note that the evolution speed \eqref{a7} exhibits a symmetric behavior with respect to  the state space geometry having the dumbbell-shape structure. This makes perfect sense because the resulting phase space corresponds exactly to the relevant quantum phase space, on which the dynamics of the $N$ spin-$1/2$ system is well established.

\subsection{Resolution of the quantum brachistochrone problem}
 To address this dynamic issue, we should determine the shortest duration possible necessary to carry out the time-optimal evolution of the considered system \eqref{c}. To accomplish this, one begins by maximizing the evolution velocity \eqref{a7} by solving the equation $d\mathrm{V}/d\theta=0$, which yields
 		\begin{equation}
 			N(N-1)(N-1-(2N-3)\sin^2\theta)\sin 2\theta=0,
 		\end{equation}
 		which entails that
 		\begin{equation}
 		  \sin\theta_{\max}=
 		     \sqrt{\frac{N-1}{2N-3}}.
 		\end{equation}
Therefore, the highest speed that the $N$ spin-$1/2$ system can achieve reads as
 		\begin{equation}
 			\mathrm{V}_{\max}=\mathtt{J}(N-1)\sqrt{\frac{N(N-1)}{8(2N-3)}},
 		\end{equation}
 		In this way, we manage to establish the maximum velocity, which depends only on the particle number making up the system. Let us now investigate the geodesic distance that the system traverses between the departure state \eqref{b} and the arrival state \eqref{c}. For this, utilizing the equation \eqref{a5}, one finds
 		\begin{equation}\label{a8}
  \mathtt{S}=\frac{{\xi}}{2}\sqrt{N(N-1) \sin^2 \theta\left[N-1-\left(N-\frac{3}{2}\right) \sin^2 \theta\right]}.
 \end{equation}
As the evolution rapidity \eqref{a7} is time-independent, the Fubini-Study distance \eqref{a8} traveled by the system has a linear behavior with time. Additionally, we note that, at each instant, the evolution speed and the distance have a similar behavior against the physical parameters $(\theta,\mathtt{J},N)$ of the system. Specifically, at singular points $\theta=0,\pi$, the distance vanishes, which is reasonable because at these points the state of $N$ spin-$1/2$ \eqref{c} is not defined. Thereby, one concludes that the Fubini-Study distance \eqref{a8} exhibits a local minimum at the point $\theta=\pi/2$, it given by 
 	\begin{equation}\label{a9}
  \mathtt{S}_{\min}=\frac{{\xi}}{2}\sqrt{\frac{N(N-1)}{2}}.
 \end{equation}	
 Thus, the minimum feasible time required for the system to conduct any quantum evolution writes
 \begin{equation}\label{a10}
 \mathrm{t}_{\min}=\frac{\mathtt{S}_{\min}}{\mathrm{V}_{\max}}=\frac{t}{(N-1)} \sqrt{2N-3},\quad \text{ with} \, N>1.
 \end{equation}
 This is the shortest duration needed to achieve a time-optimal evolution over the state circle \eqref{a4}. In specific terms, the result \eqref{a10} defines the optimal evolution condition, which is  typified by the fastest evolution and the shortest path joining the initial and final states. Therefore, the optimal evolution states can be produced through the following unitary transformation
 \begin{equation}
 		\left|\Psi_i\right\rangle \rightarrow|\Psi(\mathrm{t}_{\min})\rangle=e^{-i \mathrm{H} \mathrm{t}_{\min}}\left|\Psi_i\right\rangle.
 		\end{equation}
 		so that the set of these states makes up an optimal state circle described by the Fubini-Study metric of the form 
 		\begin{small}
 	\begin{equation}\label{a11}
 	d\mathtt{S}^2_{\mathrm{op}}=\frac{1}{4}N(N-1) \sin^2 \theta\left[N-1-\left(N-\frac{3}{2}\right) \sin ^2 \theta\right]d\xi^2_{\min}.
 		\end{equation}	
 		\end{small}
with $\xi_{\min}=\mathtt{J} \mathrm{t}_{\min}$. On other hand, we remark that the temporal condition \eqref{a10} is solely influenced by the particle number, this implies that the state circle topology affects the optimal evolution time as well. It is also proportional to the ordinary time $t$ (i.e., matches the evolution over the state circle \eqref{a4}) with a positive proportionality factor, meaning that the optimal and ordinary times have the same monotonicity. Particularly, one finds that for $N=2$ (i.e., two-spin-$1/2$ system) these two types of time coincide $(\mathrm{t}_{\min}=t)$, while for $N\ge3$ (i.e., $N$ spin-$1/2$ system), we discover that the optimal time \eqref{a10} is strictly lower than the ordinary time, and therefore the time-optimal evolution is achievable. Besides, in the thermodynamic limit $(N\to\infty)$, the optimal time decreases to zero $(\mathrm{t}_{\min}\to 0)$. In this respect, the optimal state circle \eqref{a11} coincides with a straight line since its radius $\sqrt{\mathrm{g_{\xi_{\min}\xi_{\min}}}}$ becomes infinite. As a result, we infer that the particle number and the ordinary time are two physical magnitudes exploitable for performing the time-optimal evolutions in such integrable systems. At the end of this section, we note that it is intriguing to relate the geometric and dynamic structures that we explored above with quantum entanglement, as a physical resource of great relevance in quantum information tasks. The next section focuses on this subject.
\section{Geometric and dynamic pictures of the entanglement for two-spin system $(N=2)$}\label{sec5}
 In this section, we shall study the quantum entanglement exchanged between two interacting spins under Ising model through two different perspectives; the first is geometric in nature and investigates the entanglement impact on derived geometric features such as the Fubini-Study metric, G-curvature, and the geometric phase under arbitrary and cyclic evolutions. The second is dynamic in nature and explores the entanglement effect on the evolution speed as well as the Fubini-Study distance covered by the system. Importantly, we address the quantum Brachistochrone problem depending on the entanglement degree.
\subsection{Entanglement degree of the two-spin system} 	
The wave function of the global quantum system \eqref{c} is narrowed for a two-spin system to the form
 \begin{align}\label{a12}
 		|\Psi(t)\rangle=& e^{-i\xi(t)}\cos^{2}
 		\frac{\theta}{2}
 		 |\uparrow\uparrow\rangle+\frac{1}{2} e^{i\varphi}\sin\theta (|\uparrow\downarrow\rangle+|\downarrow\uparrow\rangle)\notag\\&+e^{i(2\varphi-\xi(t) )} \sin^2\frac{\theta}{2}
 		|\downarrow\downarrow\rangle. 
 		\end{align} 	
Hence, the two-spin state space, on which the dynamics of the system takes place, is defined by the following metric tensor	
 \begin{equation}\label{a14}
 d \mathtt{S}^2=\frac{1}{2} d \theta^2+\frac{1}{4} \sin ^2 \theta\left(2-\sin ^2 \theta\right) d \xi^2.
\end{equation} 			
 Using the Wootters concurrence expression given in Ref. \cite{Wootters1998}, one obtains, after a simple calculation, the entanglement amount contained in the two-spin state \eqref{a12} of the form	
 		\begin{equation}\label{a13}
 			\boldsymbol{\mathscr{C}}=\sin^2\theta |\sin\xi|.
 		\end{equation}  
It is the same for any two spins of the entire system \eqref{c}. In other terms, each spin pair is quantum-correlated as much as any other pair. Interestingly, we observe that the two-spin entanglement \eqref{a13} evolves periodically with time, signifying that it is impacted by the dynamics of the system. Moreover, it relies on the initial parameter $\theta$, showing that the entanglement degree  is also governed by the starting state choice. Notice that for $\xi=\pi/2$ and $\theta=\pi/2$, the two-spin state \eqref{a12} reaches its maximum entanglement value $(\boldsymbol{\mathscr{C}}=1)$, whereas for $\theta=0$ or $\pi$, the two spins  can never be entangled $(\boldsymbol{\mathscr{C}}=0)$, because the corresponding initial states $|\Psi_i\rangle=|\uparrow\uparrow\rangle$ or $|\downarrow\downarrow\rangle$ are the Hamiltonian eigenstates. This can be also justified topologically and geometrically by the existence of a conical defect close to these two singular points.
\subsection{Geometrical picture of the entanglement} 		
In order to evoke the geometric aspect of the quantum correlations between the two spins under study, we suggest a thorough description illustrating the nexus between the entanglement and the geometric structures derived above. Setting the equation \eqref{a13} into \eqref{a14}, we give the Fubini-Study metric identifying the two-spin state space in terms of the concurrence as
\begin{widetext}
\begin{equation}\label{a15}
d{\mathtt{S}^2} = \frac{{d{\boldsymbol{\mathscr{C}}^2}}}{{8\boldsymbol{\mathscr{C}}(\left| {\sin \xi } \right| - \boldsymbol{\mathscr{C}})}} - \frac{{d\boldsymbol{\mathscr{C}}d\xi }}{{4\tan \xi (\left| {\sin \xi } \right| - \boldsymbol{\mathscr{C}})}} + \frac{\boldsymbol{\mathscr{C}}}{4}\left( {\frac{1}{{2{{\tan }^2}\xi (\left| {\sin \xi } \right| - \boldsymbol{\mathscr{C}})}} + \frac{{2\left| {\sin \xi } \right| - \boldsymbol{\mathscr{C}}}}{{{{\sin }^2}\xi }}} \right)d{\xi ^2}.
\end{equation}
\end{widetext}
which can be transformed in its diagonal form
\begin{equation}
 d \mathtt{S}^2=\frac{1}{8 \boldsymbol{\mathscr{C}}_r\left(1-\boldsymbol{\mathscr{C}}_r\right)} d \boldsymbol{\mathscr{C}}_r^{ 2}+\frac{1}{4} \boldsymbol{\mathscr{C}}_r\left(2-\boldsymbol{\mathscr{C}}_r\right) d \xi^2,
\end{equation}
with $\boldsymbol{\mathscr{C}}_r={\boldsymbol{\mathscr{C}}}/{|\sin\xi|}$ denotes the reduced concurrence varying in the interval $[0,1]$. Thereby, we have managed to reparameterize the relevant phase space \eqref{a14} according to the amount of entanglement shared between the two spins and the evolution time, which are two measurable physical magnitudes. This demonstrates the feasibility of examining, experimentally, all the geometrical, topological and dynamical characteristics of this state space, namely the phase space geometry, the quantum phases, the evolution speed, and the geodesic distance covered by the two-spin system \eqref{a12} during its evolution. Importantly, the quantum entanglement serves in shrinking the state space dimension. For instance, the states of two spins with the same  entanglement level $(\text{i.e.,} \,\boldsymbol{\mathscr{C}}=\text{constant})$ are located on closed one-dimensional manifolds defined by
\begin{equation}
d{\mathtt{S}^2} =  \frac{\boldsymbol{\mathscr{C}}}{4}\left( {\frac{1}{{2{{\tan }^2}\xi (\left| {\sin \xi } \right| - \boldsymbol{\mathscr{C}})}} + \frac{{2\left| {\sin \xi } \right| - \boldsymbol{\mathscr{C}}}}{{{{\sin }^2}\xi }}} \right)d{\xi ^2}.
\end{equation}
They are, in fact, closed curves, along the metric component $\mathrm{g}_{\xi\xi}$, on the whole state space \eqref{a15}. On the other hand, the states with the same degree of reduced entanglement $(\text{i.e.,} \,\boldsymbol{\mathscr{C}}_r=\text{constant})$ are located on circles identified by
\begin{equation}
 d \mathtt{S}^2=\frac{1}{4} \boldsymbol{\mathscr{C}}_r\left(2-\boldsymbol{\mathscr{C}}_r\right) d \xi^2,
\end{equation}
whose radii $\mathtt{R}=\sqrt{ \boldsymbol{\mathscr{C}}_r\left(2-\boldsymbol{\mathscr{C}}_r\right)}/2$ depend on the level of reduced entanglement taken into account. In this way, we demonstrate the relevance of utilizing quantum correlations to minimize the dimensionality of the state space \eqref{a15}. This finding is applicable to all integrable quantum systems.\par
\begin{figure}[h]
\begin{center}
\includegraphics[scale=0.48]{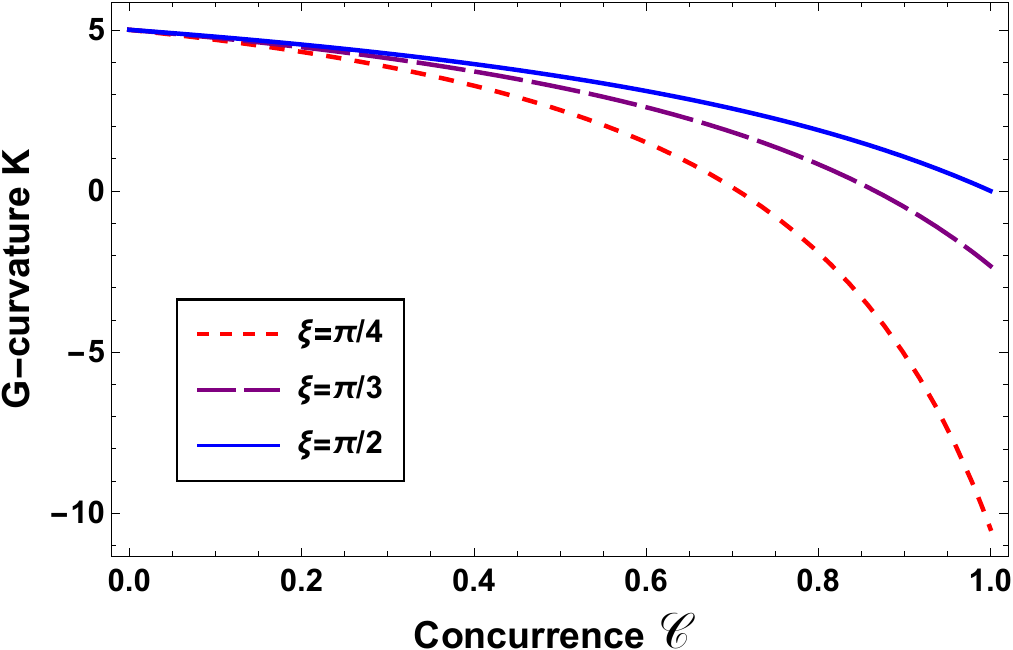}
\caption{The G-curvature \eqref{a16} versus the concurrence \eqref{a13} for some values of $\xi$.}\label{amk}
\end{center}
\end{figure}

In the same framework, we can also investigate the influence of entanglement on the G-curvature of the two-spin phase space \eqref{a15}.  As a matter of fact, putting the equation \eqref{a13} into \eqref{h}, we derive the G-curvature according to the concurrence under the form
\begin{equation}\label{a16}
\mathrm{K} = 4\left[ {2 + \frac{{\left| {\sin \xi } \right|\left( {\boldsymbol{\mathscr{C}} - 3\left| {\sin \xi } \right|} \right)}}{{{{\left( {\boldsymbol{\mathscr{C}} - 2\left| {\sin \xi } \right|} \right)}^2}}}} \right].
\end{equation}
This outcome demonstrates again the explicit reliance of the state space geometry on the entanglement amount exchanged between the two interacting spins. From the equation \eqref{a16}, we observe that for $\xi=0$ (i.e., no evolution), the G-curvature is $\boldsymbol{\mathscr{C}}$-independent and takes a constant value $(\mathrm{K}=8)$, which corresponds to the curvature of the initial state sphere \eqref{a17}, while for $\xi>0$ (i.e., evolution case) it is $\boldsymbol{\mathscr{C}}$-dependent and its behavior versus the entanglement is shown in the figure \eqref{amk}. We notice that the G-curvature diminishes as the entanglement amount exchanged between the two spins increases. This can be explained by the fact that the existence of quantum correlations causes a decrease in the state space curvature. Further, the G-curvature reaches negative values for the entanglement degrees verifying the condition 
\begin{equation}\label{a17}
{\left| {\sin \xi } \right|\left( {\boldsymbol{\mathscr{C}} - 3\left| {\sin \xi } \right|} \right)}<-2{{{\left( {\boldsymbol{\mathscr{C}} - 2\left| {\sin \xi } \right|} \right)}^2}}.
\end{equation}
This elucidates the quantum correlation impact between the two particles in the compactification of the related state space \eqref{a15} when the requirement \eqref{a17} is met. It is interesting to note that the separable states $(\boldsymbol{\mathscr{C}}=0)$ are housed in the regions of highest curvature $\mathrm{K}_{\max}=5$,
 \noindent  whereas the states of maximum entanglement $(\boldsymbol{\mathscr{C}}=1)$ are housed in the regions of lowest curvature
\begin{equation}\label{a18}
{{\mathrm{K}}_{\min }} = 4\left[ {2 - \frac{{\left| {\sin \xi } \right|\left( {3\left| {\sin \xi } \right| - 1} \right)}}{{{{\left( {2\left| {\sin \xi } \right| - 1} \right)}^2}}}} \right].
\end{equation}
Thus, we conclude that the information about the entanglement degree of the two-spin system \eqref{a12} allows to determine its localization on the corresponding phase space \eqref{a15}. This clearly proves the deterministic character of geometric quantum mechanics, this latter is based on the geometrization of Hilbert space by introducing the concept of quantum phase space analogous to that of classical mechanics \cite{Brody2001,Zhang1995}.\par
\begin{figure}[h]
\begin{center}
\includegraphics[scale=0.48]{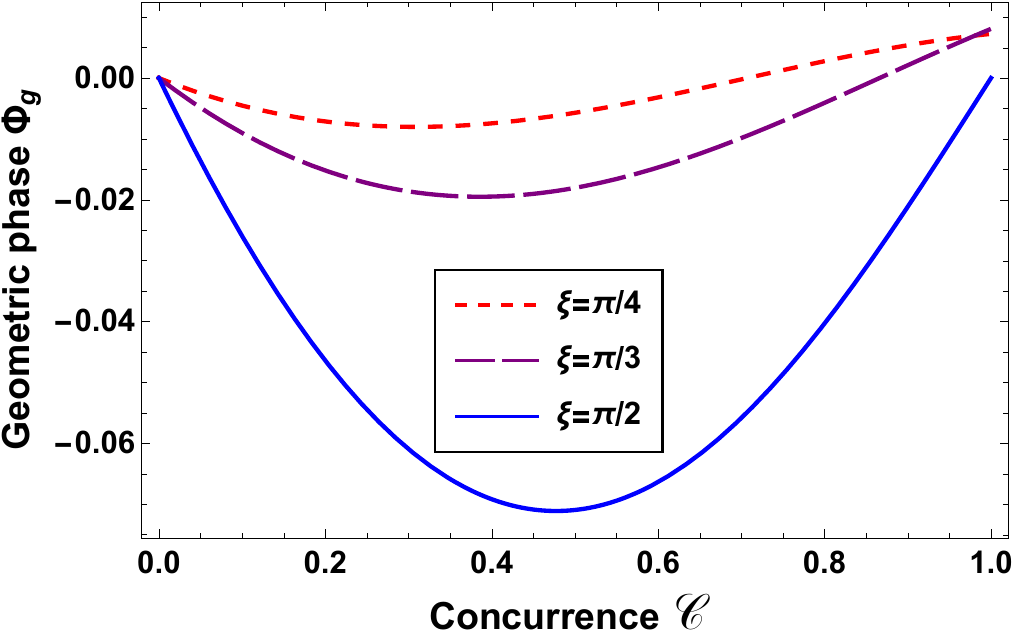}
\caption{The geometric phase \eqref{a19} versus the concurrence \eqref{a13} for some values of $\xi$.}\label{amkp}
\end{center}
\end{figure}

The connection between the geometric phase and the quantum entanglement can be also explored here. Indeed, reporting the equation \eqref{a13} into \eqref{p}, we obtain the geometric phase gained by the two-spin state \eqref{a12}  in terms of the concurrence as
\begin{small}
\begin{equation}\label{a19}
  \Phi_{\operatorname{g}}=-\arctan\left[
  \dfrac{(2\left| {\sin \xi } \right|-\boldsymbol{\mathscr{C}})\sin\xi}{(2\left| {\sin \xi } \right|-\boldsymbol{\mathscr{C}})\cos\xi+\boldsymbol{\mathscr{C}}}
  \right]+\xi\begin{pmatrix}
  1-\dfrac{\boldsymbol{\mathscr{C}}}{2\left| {\sin \xi } \right|}
  \end{pmatrix}.
  \end{equation}
\end{small}
Notice that the geometric phase is determined by the two new physical degrees of freedom : entanglement and time. This means that it depends on each point (i.e., system physical state) of the underlying phase space \eqref{a15}. Consequently, we can say that the geometric phase depends both on the path followed by the system and the geometry of the state space. Since the geometrical phase is defined in terms of these two measurable physical magnitudes (i.e., $\boldsymbol{\mathscr{C}}$ and $\xi$), this offers us the ability to measure it experimentally for any arbitrary evolution process of the system. This result is extremely important because it can be exploited to build efficient quantum circuits based on the amount of entanglement exchanged between the two interacting spins. To further highlight the interplay between the geometric phase and entanglement, we have graphed the reliance of the Eq. \eqref{a19} with respect to the concurrence for some values of $\xi$ in Figure \eqref{amkp}, we observe that the geometric phase \eqref{a19} gained by the two-spin system \eqref{a12} during its evolution from the separable state $(\boldsymbol{\mathscr{C}}=0)$ to the maximum entanglement state $(\boldsymbol{\mathscr{C}}=1)$ exhibits approximately a parabolic behavior. From here, we can divide its evolvement into two main stages : the first stage involves the geometric phase decrease along the concurrence interval  $\boldsymbol{\mathscr{C}}\in[0,\boldsymbol{\mathscr{C}}_{\text{c}}]$ with $\boldsymbol{\mathscr{C}}_{\text{c}}$ denotes the  critical entanglement degree in which this phase reaches its minimum value (see the figure \eqref{amkp}), it is given explicitly as
\begin{equation}
\boldsymbol{\mathscr{C}}_{\text{c}}=\sin \xi  - \cot \frac{\xi }{2}\sqrt {\frac{{\sin \xi }}{\xi }\left( {2 - \xi \sin \xi  - 2\cos \xi } \right)}. 
\end{equation}
In this stage, the evolving state \eqref{a12} acquires a negative geometric phase, which can be interpreted as the geometric phase part lost by the system. Geometrically, we can say that during parallel transport, the state vector \eqref{a12} rotates clockwise (i.e., makes an angle of negative sign) with respect to the separable state (i.e., starting state). Thereby, we discover that, in the region $[0,\boldsymbol{\mathscr{C}}_{\text{c}}]$, the quantum correlations favors the loss of the geometric phase. The second stage concerns the geometric phase increase along the interval $\boldsymbol{\mathscr{C}}\in [\boldsymbol{\mathscr{C}}_{\text{c}},1]$ (i.e., reverse behavior), the evolving state \eqref{a12} accumulates a positive geometric phase, which can be viewed as the geometric phase part gained by the system. We can say, geometrically, that during parallel transport, the state vector \eqref{a12} rotates counterclockwise (i.e., makes an angle of positive sign) with respect to the separable state. In this way, we find that, in the region $ [\boldsymbol{\mathscr{C}}_{\text{c}},1]$, the quantum correlations favors the gain of the geometric phase. Accordingly, the geometric phase behavior versus the entanglement is approximately symmetric with respect to the critical value $\boldsymbol{\mathscr{C}}_{\text{c}}$, this is mainly due to the dumbbell-shape structure of the underlying phase space \eqref{a15}. On the practical side, the quantum entanglement is then an interesting physical resource that can be exploited experimentally to control the geometric phase resulting from the evolution processes of the two-spin system.\par
\begin{figure}[h]
\begin{center}
\includegraphics[scale=0.48]{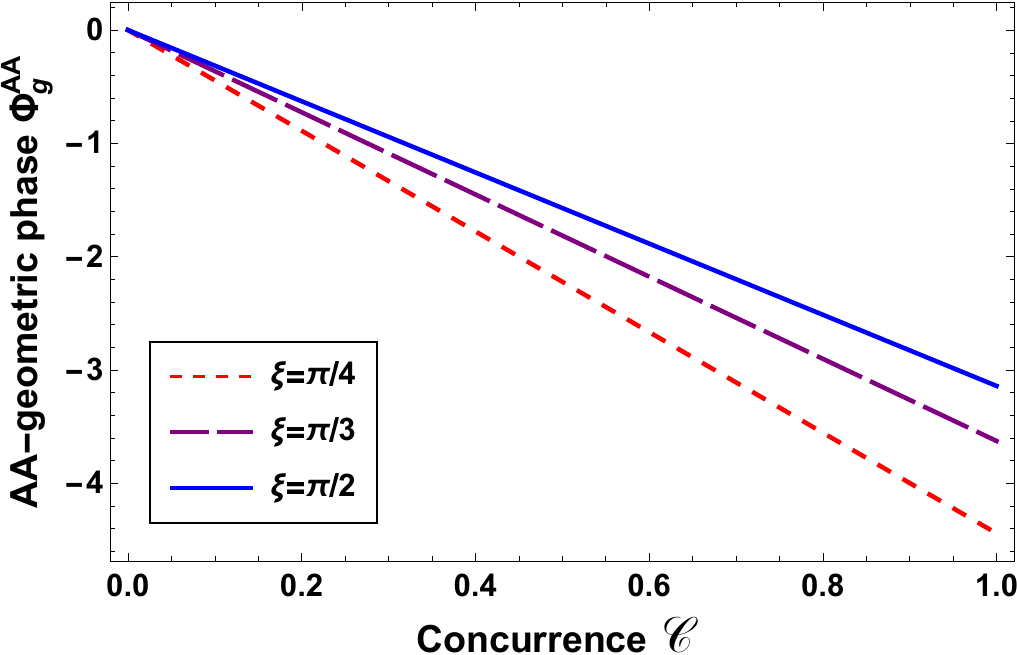}
\caption{The AA-geometric phase \eqref{a20} versus the concurrence \eqref{a13} for some values of $\xi$.}\label{amkr}
\end{center}
\end{figure}

In the cyclic evolution scenario, the geometric phase can be also investigated in connection to the entanglement. Indeed, inserting the equation \eqref{a13} into \eqref{x1}, we give the AA-geometric phase accumulated by the evolved state \eqref{a12} in relation to the concurrence as
\begin{equation}\label{a20}
\Phi_{\operatorname{g}}^{\mathrm{AA}}= -{\pi}\frac{\boldsymbol{\mathscr{C}}}{\left| {\sin \xi } \right|}.
\end{equation}
It is proportional to the entanglement level between the two spins with a negative proportionality factor, implying that the AA-geometric phase decreases linearly as entanglement increases. For the sake of clarity, we display this behavior in Figure \eqref{amkr}, we observe indeed that the more the system is entangled, the more it accumulates an AA-geometric phase of negative sign.
 This is roughly the same behavior as we observed for the geometric phase \eqref{a19} in the first stage (i.e., in the region $[0,\boldsymbol{\mathscr{C}}_{\text{c}}]$), and hence the same interpretations can be provided for the AA-geometric phase. Regarding the topological phase resulting from the cyclic evolutions of the two-spin system, it is given by $\Phi_{\operatorname{top}}^{\mathrm{AA}} =-2\pi$. Thus, it is unaffected by the entanglement, because it constitutes the AA-geometrical phase part receiving no contribution from the dynamic part.
\subsection{Dynamical picture of the entanglement} 
\begin{figure}[h]
\begin{center}
\includegraphics[scale=0.48]{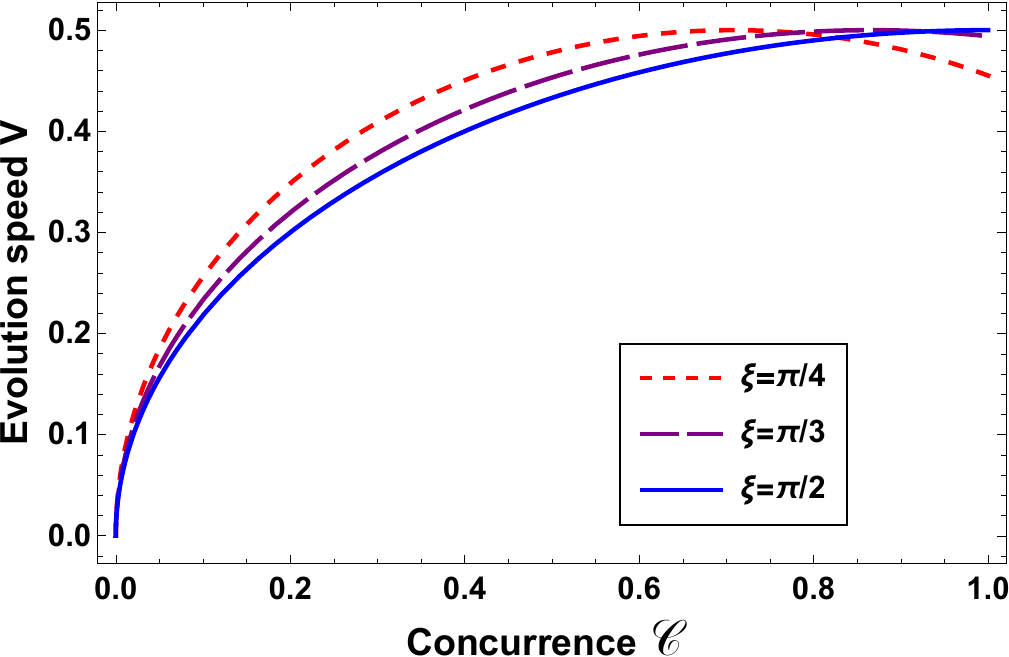}
\caption{The evolution speed \eqref{a21} verus the concurrence \eqref{a13} for some values of $\xi$ with $\mathtt{J}=1$.}\label{amkl}
\end{center}
\end{figure}

To close this section, it would be interesting to examine the dynamical aspect of entanglement by highlighting the link between the entanglement amount exchanged between the two spins and the relevant dynamical properties, such as the evolution speed and the traveled geodesic distance during a given evolution process over the resulting phase space \eqref{a15}. As a result, we address the quantum brachistochrone problem by relying on the entanglement level of the two-spin system. To accomplish this, using the equation \eqref{a13} into \eqref{a7}, the related evolution speed can be expressed in relation to the concurrence as follows
 		\begin{equation}\label{a21}
 			\mathrm{V}=\frac{\mathtt{J}}{2\left|\sin\xi\right|}\sqrt{{\boldsymbol{\mathscr{C}}}\begin{pmatrix}
 				2\left|\sin\xi\right|-{\boldsymbol{\mathscr{C}}}
 				\end{pmatrix}}.
 		\end{equation}
In this way, we manage to relate the rapidity of the two-spin system with its entanglement degree. In other words, the result \eqref{a21} reflects the explicit relatedness between the system dynamics and the evolvement of the quantum correlations. Consequently, we infer that the dynamical characteristics of such a system are determinable through its entanglement state. The evolution speed reliance according to the concurrence is shown in Figure \eqref{amkl}.\par
 Interestingly, the variation of the evolution velocity \eqref{a21} is split into two different parts : the first part shows the  the evolution velocity increase of the two-spin state until it attains its highest value $\mathrm{V}_{\max}={\mathtt{J}}/2$, matching the critical entanglement level $\boldsymbol{\mathscr{C}}=\boldsymbol{\mathscr{C}}_{\text{c}}^\prime=\left|\sin\xi\right|$. This 
 proves that, in this part, the existence of quantum correlations speed up the system evolution over the related phase space \eqref{a15}.  The second part concerns the concurrence interval $\boldsymbol{\mathscr{C}}\in[\boldsymbol{\mathscr{C}}_{\text{c}}^\prime, 1]$, in which the evolution speed has reversed its monotonicity, it diminishes continuously until it achieves its local minimum $\mathrm{V}(\boldsymbol{\mathscr{C}}=1)$. This signifies that, in this second part, the quantum correlations slow down this evolution. As a result, we conclude that the system dynamics is controllable by its entanglement level. This outcome can be usefully exploited in quantum information protocols. Utilizing the equation \eqref{a5}, we can also establish the Fubini-Study distance traveled by the two-spin state \eqref{a12} in relation to the concurrence, it is found as
 	\begin{equation}\label{a22}
 			\mathtt{S}=\frac{\xi}{2\left|\sin\xi\right|}\sqrt{{\boldsymbol{\mathscr{C}}}\begin{pmatrix}
 				2\left|\sin\xi\right|-{\boldsymbol{\mathscr{C}}}
 				\end{pmatrix}}.
 		\end{equation}
 Thereby, we arrive at expressing the Fubini-Study distance (i.e., dynamical observable), relating any two quantum states over the two-spin phase space \eqref{a15}, in terms of the entanglement level and the evolution time. This proves again the feasibility to investigate experimentally the dynamical properties, which will motivate their exploitation in the novel applications of quantum technology \cite{Sato2012,KirdiS2023}.\par		
 		\begin{figure}[h]
\begin{center}
\includegraphics[scale=0.48]{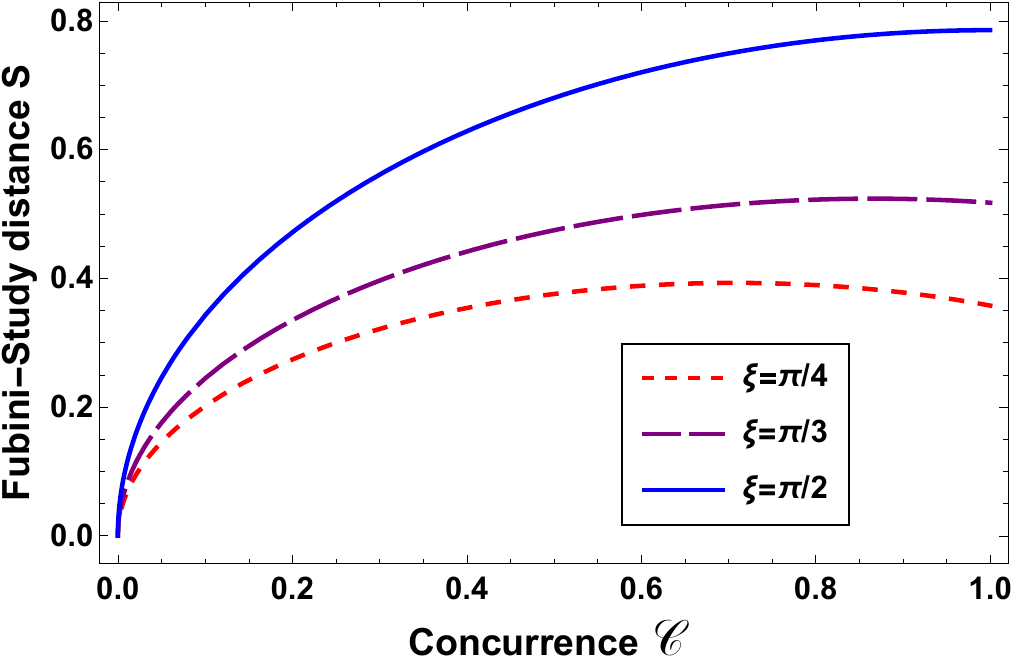}
\caption{The Fubini-Study distance \eqref{a22} versus the concurrence \eqref{a13} for some values of $\xi$.}\label{amkn}
\end{center}
\end{figure}
\begin{figure}[h]
\begin{center}
\includegraphics[scale=0.48]{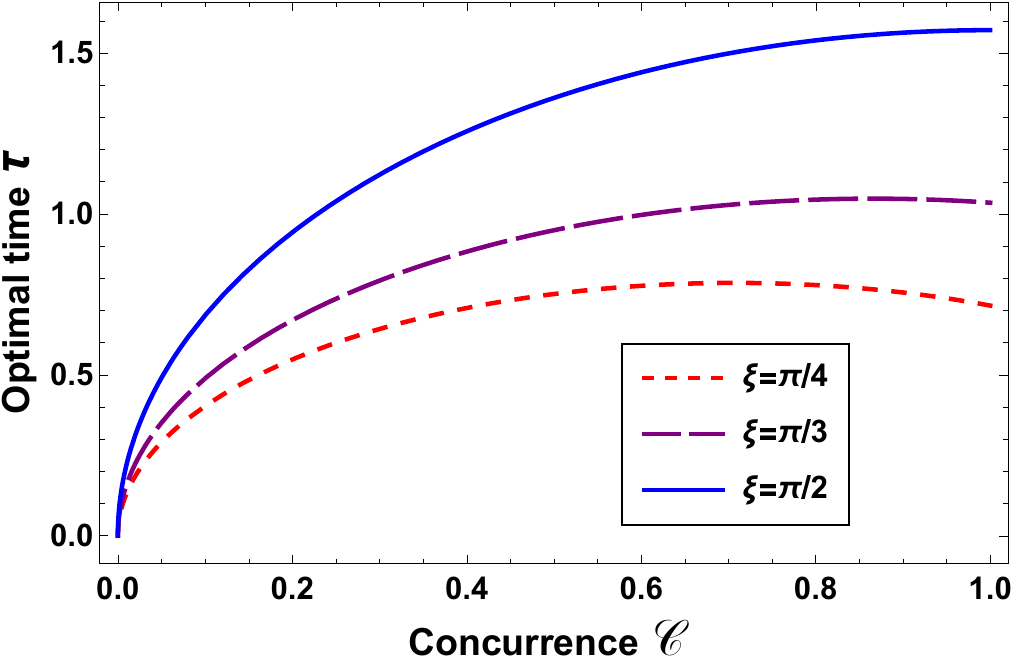}
\caption{The optimal time \eqref{a23} versus the concurrence \eqref{a13} for some values of $\xi$ with $\mathtt{J}=1$.}\label{amkm}
\end{center}
\end{figure}

	
By scrutinizing the two figures \eqref{amkl} and \eqref{amkn}, we find that the Fubini-Study distance \eqref{a22} exhibits, with respect to the entanglement degree, the same behavior as the evolution velocity \eqref{a21}, and thus the same conclusions can be achieved. Let us now solve the quantum brachistochrone issue for the two interacting spins based on their entanglement.  To realize this, maximizing the evolution speed \eqref{a21} with respect to the concurrence, the shortest time required to achieve a time-optimal evolution, over the relevant phase space \eqref{a15}, reads as
 \begin{equation}\label{a23}
 \boldsymbol{\tau}=\frac{\mathtt{S}}{\mathrm{V}_{\max}}=\frac{\xi}{\mathtt{J}\left|\sin\xi\right|}\sqrt{{\boldsymbol{\mathscr{C}}}\begin{pmatrix}
 				2\left|\sin\xi\right|-{\boldsymbol{\mathscr{C}}}
 				\end{pmatrix}}.
\end{equation} 		
So, the optimal time \eqref{a23} depends on both the ordinary time, the coupling constant, and the entanglement level of the system. Specifically, we observe that for $\boldsymbol{\mathscr{C}}=0$  the optimal time cancels out $(\boldsymbol{\tau}=0)$, this because the evolving state \eqref{a12} coincides with the disentangled starting state $|\Psi_i\rangle=|++\rangle$ (i.e., no evolvement). For the critical entanglement level $\boldsymbol{\mathscr{C}}=\boldsymbol{\mathscr{C}}_{\text{c}}^\prime$ the optimal time attains its highest value $(\boldsymbol{\tau}=t)$, signifying that the optimal and ordinary evolutions of the two-spin system coincide, whereas for $\boldsymbol{\mathscr{C}}\in \,]0,\boldsymbol{\mathscr{C}}_{\text{c}}^\prime[\, \cup\,  ]\boldsymbol{\mathscr{C}}_{\text{c}}^\prime,1]$  the optimal time is strictly less than the ordinary time $(\boldsymbol{\tau}<t)$. In this respect, the optimal evolution states can be generated via the unitary operation given by
 \begin{equation}
 		\left|\Psi_i\right\rangle \rightarrow|\Psi(\boldsymbol{\tau})\rangle=e^{-i \mathrm{H} \boldsymbol{\tau}}\left|\Psi_i\right\rangle.
 		\end{equation} 
In fact, they make up  one-dimensional space of optimal states, over the whole space \eqref{a15}, defined by the metric tensor 
 		\begin{equation}\label{a24}
 d {{\mathtt{S}}}^2_{\text{opt}}=\frac{{\boldsymbol{\mathscr{C}}}}{4\sin^2\xi}\left(2\left|\sin\xi\right|-\boldsymbol{\mathscr{C}}\right) d\boldsymbol{\xi}^2,
\end{equation}
with $\boldsymbol{\xi}=\mathtt{J}\boldsymbol{\tau}$. The behavior of the optimal time \eqref{a23} according to the entanglement is illustrated in Figure \eqref{amkm}, we discover that  the
lower the entanglement level (resp. the ordinary time) of 
the two spins  ${\boldsymbol{\mathscr{C}}}\to 0$ (resp. $t\to 0$), the shorter the optimal time $\boldsymbol{\tau}\to 0$. Accordingly, we conclude that the entanglement and the ordinary time are two physical magnitudes exploitable for realizing the optimal evolutions in such integrable systems. These kinds of evolutions are of paramount importance in quantum computing for designing good quantum algorithms \cite{Albornoz2019,Carlini2007}.

 \section{Conclusion}\label{sec6}		
 To summarize, we investigated a physical system consisting of $N$ interacting spin-$1/2$ under all-range Ising model. We assumed that the starting state is the tensor product of $N$ eigenstates of the spin-$1/2$ projection operator on the positive direction denoted by the unit vector. After applying the time evolution propagator, the obtained evolved state is identified by three dynamical degrees of freedom, which are the spherical angles $(\theta,\varphi)$ and the dynamical parameter $\xi$ \eqref{c}. We established the Fubini-Study metric and explored that the related quantum phase space is a closed two-dimensional manifold \eqref{f}. Moreover, by examining the G-curvature in the framework of the Gauss-Bonnet theorem, we demonstrated that this state space has both a dumbbell-shape structure and a spherical topology. The geometric phase acquired by the $N$ spin-$1/2$ system is also studied within the arbitrary and cyclic evolution processes \eqref{p}. This is achieved by evaluating the difference between the total and dynamic phases. We found that the geometric phase, in the arbitrary evolution case, varies nonlinearly with the time, reflecting its dynamic character. Geometrically, we showed that it is affected both by the evolution trajectory taken by the system and the associated state space geometry. On this view, we concluded that the geometric phase can be exploited to parameterize the possible evolution trajectories of the system. This finding could be crucial for developing robust quantum gates and, consequently, for creating effective quantum circuits essential for quantum computation. Further, in the thermodynamic limit ($N\to\infty$), the total phase cancels out and therefore the geometric and dynamic phases coincide at any moment in the evolution process. This offers the opportunity of measuring the geometric phase experimentally. In the cyclic evolution case, we have calculated the AA-geometric phase and discovered that it is independent of the system dynamics \eqref{x1}. Otherwise, it depends only on the initial state choice (i.e., the initial parameters), signifying that it is influenced by the state space geometry and not by the evolution path followed by the system. Hence the cyclic evolution paths are not parameterizable by the AA-geometric phase. We have also derived the topological phase appearing naturally during cyclic evolutions. We found that it is proportional to the square of the particle number $N^2$,  especially it takes fractional values for $N$ odd and multiples of $\pi$ for $N$ even. This signifies that the number of spins influences the topology of the state space. The evolution speed and the Fubini-Study distance separating the quantum states are also well examined \eqref{a7}. As a result, we resolved the quantum Brachistochrone problem for the $N$ spin-$1/2$ system by determining the shortest time (i.e., optimal time) required to perform an time-optimal evolution \eqref{a10}. Such evolutions play an essential role in advancing quantum circuits by reducing the effects of decoherence and ensuring the reliability of algorithms. In this perspective, we discovered that for $N=2$ (i.e., two spin$-1/2$ system) the optimal and ordinary times coincide $(\mathrm{t}_{\min}=t)$, while for $N\ge 3$ (i.e., $N$ spin-$1/2$ system), the optimal time \eqref{a10} is strictly lower than the ordinary time, and thus the time-optimal evolution is achievable. Besides, in the thermodynamic limit $(N\to\infty)$, the optimal time decreases to zero $(\mathrm{t}_{\min}\to 0)$. In this scheme, the optimal state circle coincides with a straight line since its radius becomes infinite. Thereby, we deduced that the particle number and the ordinary time are two physical magnitudes exploitable for performing the time-optimal evolutions in such solvable systems.\par
 
On other hand, by restricting the whole system to a two-spin system, i.e., two interacting spin under the Ising model, we have studied the quantum entanglement via two approaches. The first approach is of geometric nature, in which we give the Fubini-Study metric in connection to the Wootters concurrence as a quantum correlation quantifier \eqref{a15}. This outcome may be interesting for the experimental handling of the geodesic distance between entangled states and also in the state space geometry adjustment. We proved that an increase in the entanglement degree between the two spins causes a decrease in the state space curvature until it reaches negative values, showing the quantum correlation effect in the compactification of the related state space (see Fig. \eqref{amk}). Additionally, the entanglement can be used to identify the quantum states over the space of states, for example, the states of maximum entanglement are housed in the regions of lowest curvature, whereas the separable states are housed in the regions of highest curvature. The geometric phase acquired by the two-spin system is sufficiently discussed in relation to the entanglement \eqref{a19}. We explored that the geometric phase (see Fig. \eqref{amkr}) exhibits two different behaviors with respect to the critical entanglement level $\boldsymbol{\mathscr{C}}_{\text{c}}$ $(\Phi_{\operatorname{g}}(\boldsymbol{\mathscr{C}}_{\text{c}})={\Phi_{\operatorname{g}}}_{\min})$: the first one is in the interval $[0, \boldsymbol{\mathscr{C}}_{\text{c}}]$, wherein the quantum correlations favors the loss of the geometric phase, while the second one is in the interval $[\boldsymbol{\mathscr{C}}_{\text{c}}, 1]$, wherein the quantum correlations favors the gain of the geometric phase. In the cyclic evolution scenario, the AA-geometric phase behaves similarly to the geometric phase in the first interval $[0, \boldsymbol{\mathscr{C}}_{\text{c}}]$. This  highlights the significance of quantum entanglement for controlling the geometric phase evolvement in such spin systems. The second approach is of dynamic nature, we linked the evolution speed with the concurrence \eqref{a21}, we observed that the speed (see Fig. \eqref{amkl} displays two different behaviors with respect to the critical entanglement level $\boldsymbol{\mathscr{C}}_{\text{c}}^\prime$ $(\mathrm{V}(\boldsymbol{\mathscr{C}}_{\text{c}}^\prime)=\mathrm{V}_{\max})$: the first one is in the interval $[0, \boldsymbol{\mathscr{C}}_{\text{c}}^\prime]$, wherein the quantum correlations speed up the system evolution over the relevant phase space, while the second one is in the interval $[ \boldsymbol{\mathscr{C}}_{\text{c}}^\prime, 1]$, wherein the quantum correlations slow down this evolution. Accordingly, we concluded that the system dynamics can be controlled by its entanglement degree, which is a physical resource that can be measured experimentally. The same behavior is noticed for the Fubini-Study distance separating the entangled states (see Fig. \eqref{amkn}). Finally, we solved the quantum barchistochrone problem based on the entanglement amount exchanged between the two spins. We inferred that the quantum entanglement and the ordinary time are two physical magnitudes exploitable to realize the time-optimal evolution in such a spin system. Thus, we were able to illustrate, to a significant extent, the connection between quantum entanglement and the geometric and dynamical characteristics characterizing the considered two-spin phase space.

\appendix

\section{}
In this appendix we demonstrate the result \eqref{k}, which determines the contribution of the angular defects in the Gauss-Bonnet theorem \eqref{j}. The quantum state manifold defined by metric tensor \eqref{f} possesses a two singular points at $\theta=0$ and $\pi$. Near these points, the manifold exhibits an angular defects having the form of cones. Remark that if we rewrite the metric tensor for these areas, since these areas are close to the singular points, we can express the metric up to the second order in $\theta$ as in \eqref{k01}. Based on these considerations, we shall derive the equation that allows us to evaluate the angular defect near a singular point. The angular defect can be identified as a difference between the full angle $2 \pi$ and the angle that the system traverses during one complete rotation (i.e., one period) around the singular point 
\begin{equation}\label{k02}
\Lambda=2 \pi-\frac{\mathtt{S}({\xi_{\max} })}{d}
\end{equation}
where $\xi_{\max} =2\pi$,  and $\mathtt{S}({\xi_{\max} })$ represents the distance that the system traverses during one complete rotation around the singular point  and $d$ is the distance between the system path and the singular point. Now, to compute $\mathtt{S}(2\pi) / d$ we employ the expression \eqref{k06} by taking $g_{\xi \xi}$ from the metric tensor \eqref{k01}. Thus, during the one complete rotation $\left(t=\xi_{\max } / \mathtt{J}\right)$ the system covers the following distance
\begin{equation}\label{k03}
\mathtt{S}_{\xi_{\max }}=\sqrt{g_{\xi \xi}} \xi_{\max }=\frac{1}{2}(N-1) \sqrt{N} \theta \xi_{\max }.
\end{equation}
Taking $g_{\theta \theta}$ from the metric \eqref{k01}, we give $d$ as follows
\begin{equation}\label{k04}
d=\sqrt{g_{\theta \theta}} \theta= \frac{1}{2}\sqrt{N} \theta .
\end{equation}
By introducing the equations \eqref{k03} and \eqref{k04} into \eqref{k02}, and considering that the manifold defined by the metric \eqref{f} possesses two angular defects, we find, after a simple calculation, the result \eqref{k}.\\

{\bf Disclosures:} The authors declare no conflicts of interest or personal relationships that could have appeared to influence this work, and no Data associated in the manuscript.

\end{document}